\documentclass[12pt,a4paper]{article}
\pdfoutput=1
\usepackage{color}
\usepackage{amssymb,amsmath,bm,bbold}
\usepackage{epsf}
\usepackage{epsfig}
\usepackage{relsize}
\usepackage[dvipsnames]{xcolor}
\usepackage[linktoc=page,bookmarks=false,colorlinks=false,linkbordercolor=RoyalBlue,citebordercolor=ForestGreen,urlbordercolor=CornflowerBlue]{hyperref}
\usepackage{latexsym,mathrsfs,dsfont}
\usepackage[normalem]{ulem} 
\usepackage[compress]{cite}
\usepackage{graphicx}
\usepackage{url}
\usepackage{booktabs}
\usepackage{float}
\usepackage{multirow}
\usepackage{changepage}
\usepackage[hypcap]{caption, subcaption}

\usepackage[titles]{tocloft}

\usepackage{tabularx,colortbl}

\usepackage{fancyhdr}
\advance \headheight by 3.0truept       

\allowdisplaybreaks[1]

\definecolor{red}{cmyk}{0,1,1,0.4}
\definecolor{darkgreen}{rgb}{0.0,0.6,0.0}
\definecolor{cDarkGrey}{RGB}{91,91,91}
\definecolor{cGrey}{RGB}{245,243,238}
\definecolor{cBlue}{RGB}{0,110,191}
\definecolor{cLightBlue}{RGB}{214,237,252}
\definecolor{cRed}{RGB}{196,0,100}
\definecolor{cLightRed}{RGB}{254,222,237}
\definecolor{cGreen}{RGB}{0,166,80}
\definecolor{cLightGreen}{RGB}{254,222,237}
\definecolor{cOrange}{RGB}{221,74,44}
\definecolor{cLightOrange}{RGB}{255,215,210}
\definecolor{cPurple}{RGB}{93,35,125}
\definecolor{cLightPurple}{RGB}{241,230,252}
\definecolor{cYellow}{RGB}{252,191,10}
\definecolor{cISSRBlue}{RGB}{0,111,174}
\definecolor{cISSRGrey}{RGB}{167,169,172}

\newcommand{\beq}{\begin{equation}}
\newcommand{\eeq}{\end{equation}}
\newcommand{\be}{\begin{equation}}
\newcommand{\ee}{\end{equation}}
\newcommand{\bi}{\begin{itemize}}
\newcommand{\ei}{\end{itemize}}
\newcommand{\ba}{\begin{array}}
\newcommand{\ea}{\end{array}}
\newcommand{\beqa}{\begin{eqnarray}}
\newcommand{\eeqa}{\end{eqnarray}}
\newcommand{\bea}{\begin{eqnarray}}
\newcommand{\eea}{\end{eqnarray}}
\newcommand{\beqn}{\begin{eqnarray}}
\newcommand{\eeqn}{\end{eqnarray}}

\newcounter{TODO}

\newcommand{\mev}{\text{MeV}}
\newcommand{\GeV}{\,\text{GeV}}

\newcommand{\vcb}{|V_{cb}|}
\newcommand{\vtd}{|V_{td}|}
\newcommand{\vub}{|V_{ub}|}
\newcommand{\vts}{|V_{ts}|}
\newcommand{\vus}{|V_{us}|}

\def\kpn{K^+\rightarrow\pi^+\nu\bar\nu}
\def\klpn{K_{L}\rightarrow\pi^0\nu\bar\nu}

\newcommand{\eps}{\epsilon}

\begin{document}

\begin{flushleft}
\end{flushleft}

\vspace{-14mm}
\begin{flushright}
  AJB-22-7
\end{flushright}

\medskip

\begin{center}
{\Large\bf\boldmath
  On the Superiority of the $\vcb-\gamma$ Plots over\\ the Unitarity Triangle
  Plots in the 2020s
}
\\[1.0cm]
{\bf
    Andrzej~J.~Buras
}\\[0.3cm]

{\small
TUM Institute for Advanced Study,
    Lichtenbergstr. 2a, D-85747 Garching, Germany \\[0.2cm]
Physik Department, TU M\"unchen, James-Franck-Stra{\ss}e, D-85748 Garching, Germany
}
\end{center}

\vskip 0.5cm

\begin{abstract}
  \noindent
  The Unitarity Triangle (UT) plots played already for three decades an important role in the tests
  of the Standard Model (SM) and the determination  of the CKM parameters.
  As of 2022, among the  four CKM parameters, $|V_{us}|$ and $\beta$ are already measured with respectable precision, while this is not the case of $|V_{cb}|$ and $\gamma$. In the case of $|V_{cb}|$ the main obstacle  are the significant
  tensions between its inclusive and exclusive determinations from tree-level decays  and it could
  still take some years before a unique value of this parameter will be known.
  The present uncertainty in  $\gamma$ of  $4^\circ$ from tree-level decays
    will be reduced to $1^\circ$ by the LHCb and Belle II collaborations in the coming years.   Unfortunately in the common UT plots $|V_{cb}|$ is not seen and
  the experimental improvements in the determination of $\gamma$
  from tree-level decays at the level of a few degrees are difficult to appreciate. In view of these   deficiencies of the UT plots with respect to $|V_{cb}|$ and $\gamma$ and the central role these two  CKM parameters will play in this decade, the recently proposed plots of $|V_{cb}|$ versus $\gamma$ extracted from various processes   appear to be superior to the UT plots in the flavour phenomenology of the 2020s.   We illustrate this idea with $\Delta F=2$ observables $\Delta M_s$, $\Delta M_d$,  $\varepsilon_K$  and with rare decays
  $B_s\to\mu^+\mu^-$, $B_d\to\mu^+\mu^-$,  $K^+\to\pi^+\nu\bar\nu$ and $K_L\to\pi^0\nu\bar\nu$.
In particular the power of  $\varepsilon_K$, $\mathcal{B}(K^+\to\pi^+\nu\bar\nu)$
and $\mathcal{B}(K_L\to\pi^0\nu\bar\nu)$ in the determination of $|V_{cb}|$, due to their
strong dependence on $|V_{cb}|$, is transparently exhibited in this manner.
Combined with future reduced errors on $\gamma$ and $|V_{cb}|$ from tree-level decays such plots can better exhibit possible inconsistencies between various determinations
of these two parameters, caused by new physics, than it is possible with the UT plots. 
 
\end{abstract}

\thispagestyle{empty}
\newpage
\setcounter{page}{1}

%
%
%
\section{Introduction}
\begin{figure}
\centering
\includegraphics[width = 0.55\textwidth]{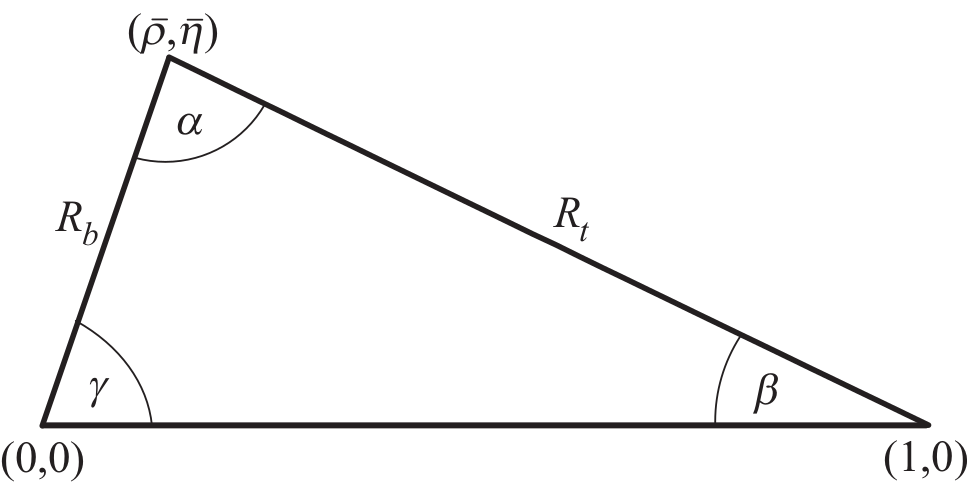}
 \caption{\it The Unitarity Triangle. }\label{UUTa}
\end{figure}

The unitary CKM matrix \cite{Cabibbo:1963yz,Kobayashi:1973fv} can be conveniently parametrized by the four
parameters
\be\label{4CKM}
\boxed{\vus,\qquad \vcb, \qquad \beta, \qquad \gamma}
\ee
with $\beta$ and $\gamma$ being two angles in the UT, shown in  Fig.~\ref{UUTa}.

By now $\vus$, $\beta$ and $\gamma$, as measured in tree-level decays,
are found to be  \cite{Zyla:2020zbs}
 \be\label{betagamma}
 \vus=0.2253(8),\qquad \beta=22.2(7)^\circ, \qquad
\gamma = (63.8^{+3.5}_{-3.7})^\circ \,,
   \ee
  with the last one being the most recent result from the LHCb \cite{LHCb:2021dcr}.
  Moreover, in the coming years the determination of $\gamma$ by the LHCb \cite{Cerri:2018ypt,Bediaga:2018lhg} and Belle II \cite{Belle-II:2018jsg} collaborations should be significantly improved with the uncertainty brought down to
  $1^\circ$. Also some reduction of the error on $\beta$
  from tree-level decays is expected. A review of various methods
  can be found in the Chapter 8 of \cite{Buras:2020xsm}.

  The situation with $\vcb$ is very different. There is a persistent tension
  between its inclusive and exclusive determinations \cite{Bordone:2021oof,Aoki:2021kgd}\footnote{The exclusive value
  for $\vcb$ should be considered as preliminary.}
\be\label{HYBRID}
\vcb_{\rm incl}=42.16(50)\times 10^{-3},\qquad 
\vcb_{\rm excl}=39.21(62)\times 10^{-3},
\ee
which is clearly disturbing. This is the case when  one makes
SM predictions for rare $K$ and $B$ decays branching ratios and for quark mixing observables like $\Delta M_s$, $\Delta M_d$ and $\varepsilon_K$. In particular
rare $K$ decay branching ratios, like the ones for $\kpn$ and $\klpn$,
exhibit $\vcb^{2.8}$ and $\vcb^4$ dependence, respectively, while $|\varepsilon_K|$ the $\vcb^{3.4}$ one \cite{Buras:2021nns}. $B$ physics observables exhibit
typically $\vcb^2$ dependence. This implies large modifications
 in the SM predictions  for the observables in question when the
inclusive values of $\vcb$ are replaced by the exclusive ones \cite{Buras:2022wpw}.

This problematic is not seen directly in the usual UT plots \cite{Bona:2007vi,Charles:2004jd}, simply  because
$\vcb$ does not enter   Fig.~\ref{UUTa} explicitly. Moreover 
the bands resulting from rare $K$ decays and $\varepsilon_K$ are bound to be broad because these observables being  very sensitive to $\vcb$ suffer from
this large parametric uncertainty. However, already in 1994, it has been pointed
out in \cite{Buras:1994rj} that
the branching ratio for $\klpn$ growing like $\vcb^4$ is a powerful
tool to determine $\vcb$ in the absence of NP 
provided its branching ratio could be measured precisely. But even a
measurement of this branching ratio with $10\%$ accuracy would determine
$\vcb$ with an error of $2.5\%$ because in this decay the hadronic uncertainties
are negligible. A similar comment applies to $\kpn$ up to long distance  charm quark contribution and  to $\varepsilon_K$, for which 
 the theoretical uncertainties have been recently reduced \cite{Brod:2019rzc}.

It is then evident that the usual UT plots are not the arena where this nice
property of Kaon processes can be used properly and in fact it becomes their Achilles tendon there. Moreover, the $\vcb$ problematic cannot be
properly monitored with the help of the UT plots as $\vcb$ is hidden in computer codes.
{In my view this is a deficiency which could be eliminated if
  the usual UT plots were accompanied by the $\vcb-\gamma$ plots
  proposed recently in collaboration with Elena Venturini in \cite{Buras:2021nns,Buras:2022nrb,Buras:2022wpw}. Indeed, as far as 
$\vcb$ and $\gamma$ are concerned the
$\vcb-\gamma$ plots are superior to the usual UT plots in three ways:}
\begin{itemize}
\item
  They exhibit $\vcb$ and its correlation with $\gamma$ determined through a given observable in the SM, allowing thereby monitoring the progress 
  on both parameters expected in the coming years. Violation of this correlation in experiment will clearly indicate new physics (NP) at work.
\item
  They utilize the strong sensitivity of rare $K$ decay  processes to $\vcb$ thereby providing
  precise determination of $\vcb$
  even with modest experimental precision on their branching ratios.
\item
  They exhibit the action of $\Delta M_s$ and of $B_s$ decays which is not possible in a UT-plot.  In particular we will illustrate this with decays  $B_s\to\mu^+\mu^-$ and $B_d\to\mu^+\mu^-$. 
\end{itemize}
It appears then that the
$\vcb-\gamma$ plots for $\Delta M_s$, $\Delta M_d$ and
$\varepsilon_K$ presented in  \cite{Buras:2021nns,Buras:2022wpw} are more useful in this context than the usual UT plots that exhibit the impact of quark mixing and rare decays in
the $(\bar\varrho,\bar\eta)$ plane  \cite{Wolfenstein:1983yz,Buras:1994ec}. The goal of the present paper is to
demonstrate the usefulness of $\vcb-\gamma$ plots
also for $\kpn$, $\klpn$, $B_s\to\mu^+\mu^-$ and $B_d\to\mu^+\mu^-$ decays.
Clearly, other decays can be considered as well, in particular the
short distance contribution to $K_S\to\mu^+\mu^-$ which has the same CKM
dependence as $\klpn$.

Despite these comments we do not claim by no means that during the RUN 3 of the
LHC and the Belle II era the UT plots should be abandoned. Indeed with improved measurements
of various observables they will exhibit the CKM  unitarity tests
\be\label{rel}
\gamma \Leftrightarrow \frac{\Delta M_d}{\Delta M_s}, \qquad \beta\Leftrightarrow \frac{\vub}{\vcb},
\ee
with $\gamma$ and $\beta$ measured in tree level non-leptonic
$B$-decays. In this sense they offer complementary tests of the SM.
{Explicit relations like the ones in (\ref{rel}) are collected in Chapters
  2 and 8 in \cite{Buras:2020xsm} and some of them will be given below.}

It should be remarked that the authors
of \cite{Altmannshofer:2021uub} performed recently  a  determination of $\vcb$
and $\vub$ from loop processes, rare decays and quark mixing, by assuming
no NP contributions to these observables. To this end they used
only well measured observables in the $B$ system and $\varepsilon_K$. This strategy has already been explored in \cite{Buras:2015qea} where
$\varepsilon_K$, $\Delta M_d$ and $\Delta M_s$  and $S_{\psi K_S}$ have been considered. This was also the case of the analyses in \cite{Buras:2021nns,Buras:2022wpw}.

Our present analysis extends the $\vcb-\gamma$ strategy, developed 
in \cite{Buras:2021nns,Buras:2022wpw}, to rare $K$ decays, not considered in
\cite{Altmannshofer:2021uub}, resurrecting some old ideas from the 
1990s \cite{Buras:1994rj,Buchalla:1994tr,Buchalla:1996fp} and improving significantly on them. In this context it should be noted that in \cite{Buchalla:1994tr}
the possibility of the determination of $\beta$ from $\kpn$ and $\klpn$, basically independently of $\vcb$ and $\gamma$, has been pointed out. As pointed out
recently in  \cite{Buras:2021nns} $\beta$ can also be determined  practically
independently of  $\vcb$ and $\gamma$ either from $\kpn$ and $|\varepsilon_K|$
or $\klpn$ and $|\varepsilon_K|$.

The determination
of $\vcb$ from $\klpn$ alone proposed in  \cite{Buras:1994rj} requires 
still the value of $\gamma$ as we will see explicitly in  Section~\ref{sec:3}.
In \cite{Buchalla:1996fp} the determination of the full CKM matrix has
been achieved with the help of the measurement of $\klpn$ and  of the angle $\alpha$ in the UT in tree-level decays. However, from the present perspective, the use of $\gamma$ is favoured over $\alpha$ because of smaller hadronic uncertainties in its tree-level determinations.

The outline of our paper is as follows. In Section~\ref{sec:2}
we recall {some useful expressions related to the CKM matrix and
  the SM formulae for the seven  observables in question that exhibit
  their dependence on the CKM parameters in (\ref{4CKM}).}
In Section~\ref{sec:3} we list useful formulae for $\vcb$
as a function of $\gamma$ and $\beta$ resulting from the {\em magnificent seven}     observables considered by us. We illustrate the application of these formulae by presenting a few examples of the $\vcb-\gamma$ plots.
A brief summary and an outlook are given in Section~\ref{sec:4}.
  
\section{Basic Formulae}\label{sec:2}
\subsection{Elements of the CKM Matrix and the UT}
{It will be useful to  recall first a number of very accurate  expressions
related to the UT shown in Fig.~\ref{UUTa} that exhibit the absence of
$\vcb$ in the usual UT plots. With $\lambda=\vus$ we have
\begin{equation}\label{2.95}
R_t =  \sqrt{(1-\bar\rho)^2 +\bar\eta^2}
=\frac{1}{\lambda} \left| \frac{V_{td}}{V_{cb}} \right|,\qquad
R_b= \sqrt{\bar\rho^2 +\bar\eta^2}
= (1-\frac{\lambda^2}{2})\frac{1}{\lambda}
\left| \frac{V_{ub}}{V_{cb}} \right|.
\end{equation}

But
\be\label{vtdvub}
\vtd=\lambda  \vcb\sin\gamma,\qquad \vts=G(\beta,\gamma)\vcb,\qquad 
\vub={\lambda\sqrt{\sigma}}  \vcb\sin\beta,
\ee
where
\be
 G(\beta,\gamma)=
1 +\frac{\lambda^2}{2}(1-2 \sin\gamma\cos\beta)\,,\qquad
\sigma = \left( \frac{1}{1- \frac{\lambda^2}{2}} \right)^2\,.
\ee
Consequently 
\be\label{RtRb}
R_t= \sin\gamma,
\qquad R_b=\sin\beta\,
\ee
so that $\vcb$ disappears. }

\subsection{The Magnificent Seven}
Explicit expressions for various observables in terms of the CKM parameters
in (\ref{4CKM}), used in our paper, are the ones from  \cite{Buras:2021nns}
modified only by adjusting some reference input as stated below.

For $\Delta M_d$ and $\Delta M_s$ they are
\bea
\label{DMD}
\Delta M_d&=&
0.5065/{\rm ps}\,\left[ 
\frac{\sqrt{\hat B_{B_d}}F_{B_d}}{210.6\mev}\right]^2
\left[\frac{S_0(x_t)}{2.307}\right]
\left[\frac{\vtd}{8.67\times10^{-3}} \right]^2 
\left[\frac{\eta_B}{0.5521}\right]\,,\\ 
\label{DMS}
\Delta M_{s}&=&
17.749/{\rm ps}\,\left[
\frac{\sqrt{\hat B_{B_s}}F_{B_s}}{256.1\mev}\right]^2
\left[\frac{S_0(x_t)}{2.307}\right]
\left[\frac{\vts}{41.9\times 10^{-3}} \right]^2
\left[\frac{\eta_B}{0.5521}\right] \,.
\eea
The value {$2.307$} in the normalization of $S_0(x_t)$ is its SM value for 
{$m_t(m_t)=162.83\GeV$.} {The central values of $\vtd$ and $\vts$ exposed here are chosen to make the overall factors in these formulae to be equal
  to the experimental values of the two observables. The reference
  values for $\sqrt{\hat B_{B_d}}F_{B_d}$ and $\sqrt{\hat B_{B_s}}F_{B_s}$
  are those from the HPQCD collaboration \cite{Dowdall:2019bea} as used by us already
  in \cite{Buras:2022wpw}. Correspondingly also the resulting reference
  values for $\vtd$ and $\vts$ agree perfectly with those quoted in \cite{Dowdall:2019bea}. {Similar results for $\Delta M_d$ and $\Delta M_s$ hadronic
    matrix elements have been obtained within the HQET sum rules in
    \cite{Kirk:2017juj} and \cite{King:2019lal}, respectively.}

   Of importance is also the mixing induced CP-asymmetry in the SM  \cite{Zyla:2020zbs}
  \be\label{SpK}
  S_{\psi K_S}=\sin 2\beta=0.699(17),
  \ee
  which implies the value of $\beta$ in (\ref{betagamma}).

Next \cite{Buras:2021nns},
  \be\label{eKapp}
  |\varepsilon_K|= {2.224 \times 10^{-3} \, \left[ \frac{\vcb}{42.6 \times 10^{-3}}\right]^{3.4} \left[\frac{\sin\gamma}{\sin 64.6^\circ}\right]^{1.67}\left[\frac{\sin \beta}{\sin 22.2^\circ}\right]^{0.87}}\,.
  \ee
The expression above provides an approximation of the exact formula of \cite{Brod:2019rzc} with an accuracy of 1.5\%, in the ranges $38< \vcb \times 10^3< 43$, $60^\circ<\gamma<75^\circ$, $20^\circ<\beta<24^\circ$.

For the rare  decays $\kpn$ and $\klpn$ we have \cite{Buras:2021nns}
\begin{align}
   { \mathcal{B}(\kpn) = {(8.59 \pm 0.30)} \times 10^{-11} \,
    \bigg[\frac{\left|V_{cb}\right|}{42.6\times 10^{-3}}\bigg]^{2.8}
    \bigg[\frac{\sin\gamma}{\sin 64.6^\circ}\bigg]^{1.39}},\label{kplusApprox}
\end{align}
\begin{align}
    \mathcal{B}(\klpn) ={(2.93 \pm 0.04)} \times 10^{-11} \,
    &{\bigg[\frac{\left|V_{cb}\right|}{42.6\times 10^{-3}}\bigg]^4
\bigg[\frac{\sin\gamma}{\sin(64.6^\circ)}\bigg]^{2}
    \bigg[\frac{\sin\beta}{\sin({22.2^\circ})}\bigg]^2},\label{k0Approx}
\end{align}
where we do not show explicitly the parametric dependence on $\lambda=\vus$ and set $\lambda=0.225$. The $3.5\%$ uncertainty in $\kpn$ is dominated by the
long distance effects in the charm contribution \cite{Isidori:2005xm}, fully negligible in $\klpn$.

Similarly for $B_{s,d}\to\mu^+\mu^-$ we have
 \cite{Bobeth:2013uxa,Beneke:2019slt}} 
\begin{equation}
  \overline{\mathcal{B}}(B_{s}\to\mu^+\mu^-)_{\rm SM} = ({3.77}\pm0.06)\times 10^{-9}
\left[\frac{F_{B_s}}{{230.3}\mev}\right]^2 \left|\frac{V_{ts}}{41.9\times 10^{-3}}\right|^2\bar R_s
\label{BRtheoRpar}
\ee
\noindent
where
\be
\label{Rs}
\bar R_s=
\left(\frac{\tau_{B_s}}{{1.515} {\rm ps}}\right)\left(\frac{{0.935}}{r(y_s)}\right)
 \left(\frac{m_t(m_t)}{162.83 \GeV}\right)^{3.02}\left(\frac{\alpha_s(M_Z)}{0.1184}\right)^{0.032} \,.
 \ee
 Here $r(y_s)$ summarizes $\Delta\Gamma_s$ effects with $r(y_s)={0.935}\pm0.007$ within the SM  \cite{DescotesGenon:2011pb,deBruyn:2012wj,deBruyn:2012wk}.

For $B_d\to\mu^+\mu^-$ we have
 \begin{equation}
  {\mathcal{B}}(B_{d}\to\mu^+\mu^-)_{\rm SM} = (1.02\pm0.02)\times 10^{-10}
\left[\frac{F_{B_d}}{{190.0}\mev}\right]^2 \left|\frac{V_{td}}{8.67\times10^{-3}}\right|^2\bar R_d \,.
\label{BRtheoRpard}
\ee
\noindent
As  to an excellent {accuracy}  $r(y_d)=1$, one has this time
\be
\label{Rd}
\bar R_d=
\left(\frac{\tau_{B_d}}{1.519 {\rm ps}}\right)
 \left(\frac{m_t(m_t)}{162.83 \GeV}\right)^{3.02}\left(\frac{\alpha_s(M_Z)}{0.1184}\right)^{0.032} \,.
\ee

{These expressions exhibit the following problem.
Even if $\beta$ can be determined precisely by measuring
$S_{\psi K_S}$, the strong dependence of all rare decay branching ratios on $\vcb$
precludes in the presence of the tensions mentioned above, a useful determination of $\gamma$ with the help of a given rare decay within the SM and its confrontation with its  tree-level measurements.

In the usual UT analyses \cite{Bona:2007vi,Charles:2004jd} this problematic can  be  solved by considering ratios like the ones in (\ref{rel}) so that $\vcb$
drops out. But then no explicit information on $\vcb$, beyond the one resulting from computer codes, is provided by the UT plots.
This deficiency is removed by accompanying  the usual UT plots by  
the  $\vcb-\gamma$ plots. They illustrate on the one hand that 
considering simultaneously $|\varepsilon_K|$, $\Delta M_d$ and $\Delta M_s$
and imposing the constraint on $\beta$ in (\ref{SpK})
one is able already now to obtain respectable determination on $\vcb$
and $\gamma$ within the SM \cite{Buras:2022wpw}. On the other hand
they make clear that the  future 
precise measurements of $\gamma$ in tree-level decays and also
experimental improvements on rare decay branching ratios that are very
sensitive to $\vcb$ 
will offer very powerful tests of the SM. In particular when the experts
also agree on the unique value of $\vcb$. Let us then have a closer look
at the  $\vcb-\gamma$ plots.}


\boldmath
\section{$\vcb-\gamma$ Plots}\label{sec:3}
\unboldmath
Using the formulae just listed we can now find $\vcb$ as a function of $\gamma$ and $\beta$ resulting separately from each of the {\em magnificent seven} observables considered by us.

\boldmath
$\underline{\Delta M_d}$
\unboldmath
\bea
\label{DMD1}
\vcb&=& 42.6\times 10^{-3}\,\left[\frac{\sin(64.6^\circ)}{\sin\gamma} \right]
\left[\frac{210.6\mev}{\sqrt{\hat B_{B_d}}F_{B_d}}\right]
\left[\frac{2.307}{S_0(x_t)}\right]^{0.5}
\left[\frac{0.5521}{\eta_B}\right]^{0.5}\,
\left[\frac{\Delta M_d}{0.5065/{\rm ps}} \right]^{0.5}.
\eea

\boldmath
$\underline{\Delta M_s}$
\unboldmath
\bea
\label{DMS1}
\vcb&=& \left[\frac{41.9\times 10^{-3}}{G(\beta,\gamma)}\right]
\left[\frac{256.1\mev}{\sqrt{\hat B_{B_s}}F_{B_s}}\right]
\left[\frac{2.307}{S_0(x_t)}\right]^{0.5}
\left[\frac{0.5521}{\eta_B}\right]^{0.5}\,
\left[\frac{\Delta M_s}{17.749/{\rm ps}} \right]^{0.5}\,.~~~~~~~~
\eea

\boldmath
$\underline{|\varepsilon_K|}$
\unboldmath
\bea
\label{epsilon}
\vcb&=&42.6\times 10^{-3}\,\left[\frac{\sin(64.6^\circ)}{\sin\gamma} \right]^{0.491}
\left[\frac{\sin(\beta)}{\sin(22.2^\circ)} \right]^{0.256}
\left[\frac{|\varepsilon_K|}{2.224\times 10^{-3}} \right]^{0.294}\,.~~~~~~~~
\eea

\boldmath
$\underline{\kpn}$
\unboldmath
\bea
\label{Kplus}
\vcb&=&42.6\times 10^{-3}\,\left[\frac{\sin(64.6^\circ)}{\sin\gamma} \right]^{0.496}
\left[\frac{\mathcal{B}(\kpn)}{(8.59 \pm 0.30) \times 10^{-11}}\right]^{0.357} \,       \,.~~~~~~~~~~~~~~~~~
\eea

\boldmath
$\underline{\klpn}$
\unboldmath
\bea
\label{KL}
\vcb&=&42.6\times 10^{-3}\,\left[\frac{\sin(64.6^\circ)}{\sin\gamma} \right]^{0.50}
\left[\frac{\sin(22.2^\circ)}{\sin(\beta)}\right]^{0.50}
\left[\frac{\mathcal{B}(\klpn)}{(2.93 \pm 0.04) \times 10^{-11}}\right]^{0.25} \,       \,.~~~~
\eea

\boldmath
$\underline{B_d\to\mu^+\mu^-}$
\unboldmath
\bea
\label{Bd}
\vcb&=& 42.6\times 10^{-3}\,\left[\frac{\sin(64.6^\circ)}{\sin\gamma}\right]
\left[\frac{190.0\mev}{F_{B_d}}\right]
\frac{1}{\sqrt{\bar R_d}}
\left[\frac{\mathcal{B}(B_{d}\to\mu^+\mu^-)}{(1.02\pm0.02)\times 10^{-10}}            \right]^{0.50}\,.
\eea

\boldmath
$\underline{B_s\to\mu^+\mu^-}$
\unboldmath
\bea
\label{Bs}
\vcb&=&\left[\frac{41.9\times 10^{-3}}{G(\beta,\gamma)}\right]                 \,
\left[\frac{230.3\mev}{F_{B_s}}\right]
\frac{1}{\sqrt{\bar R_s}}
\left[\frac{\mathcal{B}(B_{s}\to\mu^+\mu^-)}{(3.77\pm0.06)\times 10^{-9}}            \right]^{0.50}\,.~~~~~~~~~~~~~
\eea

In  \cite{Buras:2021nns,Buras:2022wpw} only
$\vcb-\gamma$ plots for $\Delta M_s$, $\Delta M_d$ and
$\varepsilon_K$ have been presented. One of such plots from  \cite{Buras:2022wpw} is shown
in Fig.~\ref{fig:5}.

An important observation should be made in this plot. The $\varepsilon_K$-band is thiner  than the one coming from
$\Delta M_d$.
This is dominantly related to the stronger
dependence of $\varepsilon_K$ on $\vcb$. This fact makes the action of 
$\varepsilon_K$ less useful in the $(\bar\varrho,\bar\eta)$ plane
than that of $\Delta M_d$ while in
the $\vcb-\gamma$ plane it is reversed. The formulae above
demonstrate this in explicit terms. Moreover, while the action of $\Delta M_s$
is invisible in a UT-plot, it is clearly exhibited in  Fig.~\ref{fig:5}.

Before presenting similar plots for rare decays $\kpn$, $\klpn$,
$B_{s}\to\mu^+\mu^-$ and $B_{d}\to\mu^+\mu^-$, let us recall the SM predictions for them
\cite{Buras:2021nns,Buras:2022wpw}. These are \cite{Buras:2021nns}
\be\label{BV}
\boxed{\mathcal{B}(\kpn)_\text{SM}= {(8.60\pm 0.42)}\times 10^{-11}\,,\quad
\mathcal{B}(\klpn)_\text{SM}={(2.94\pm 0.15)}\times 10^{-11}\,,}
\ee
and \cite{Buras:2022wpw}
\be\label{LHCbTH}
\boxed{\overline{\mathcal{B}}(B_{s}\to\mu^+\mu^-)_{\rm SM} = (3.78^{+ 0.15}_{-0.10})\times 10^{-9},\quad \mathcal{B}(B_{d}\to\mu^+\mu^-)_{\rm SM} = (1.02^{+ 0.05}_{-0.03})\ \times 10^{-10}.}
\ee
Most interesting until recently was the SM prediction for the $B_{s}\to\mu^+\mu^-$
branching ratio that exhibited a $2.7\sigma$ anomaly when confronted with its experimental
value \cite{LHCb:2021awg,CMS:2020rox,ATLAS:2020acx}
\be\label{LHCbEXP1}
\overline{\mathcal{B}}(B_{s}\to\mu^+\mu^-)_{\rm EXP} = 2.86(33)\times 10^{-9}\,.
\ee
It updated the previous prediction of \cite{Bobeth:2021cxm}, based on different hadronic matrix elements,  that exhibited a $2.1\sigma$ anomaly.

However, after the most recent result from CMS collaboration, which finds
for this branching ratio $3.82(42)\times 10^{-9}$, in perfect agreement with
the SM value in (\ref{LHCbTH}), the present world average reads
\be\label{LHCbEXP2}
\overline{\mathcal{B}}(B_{s}\to\mu^+\mu^-)_{\rm EXP} = 3.45(29)\times 10^{-9}\,,
\ee
so that there is no anomaly here.

In any case, these are the most precise SM predictions for these decays to date. For $\kpn$ and $\klpn$
they were obtained by using the experimental values of  $\varepsilon_K$ and $S_{\psi K_S}$.
The ones for $B_{s,d}\to \mu^+\mu^-$ using the strategy of  \cite{Buras:2003td}
and experimental values of $\Delta M_{s,d}$\footnote{This is the  reason  why the central values in   (\ref{LHCbTH}) agree perfectly with those obtained in
 \cite{Dowdall:2019bea} where this strategy has been used as well.}.
No information on $\vcb$ was required to obtain these results and
the left-over $\gamma$ dependence in rare $K$ decay branching ratios, once $\varepsilon_K$ constraint was imposed, turned out to be negligible in the full
range $60^\circ\le\gamma\le 75^\circ$ investigated by us. One can verify all
these nice properties using the formulae above or inspect the plots in Fig.~15
of \cite{Buras:2021nns}.

 The goal of this strategy is {\em not} to   make an overall SM fit but to predict the SM branching ratios in question. 
  Yet, it also turns out  that the simultaneous description of the data for $\Delta M_d$, $\Delta M_s$,  $\varepsilon_K$ and $S_{\psi K_S}$ can be made without any participation of NP which gives an additional support for the SM predictions in (\ref{BV}) and (\ref{LHCbTH}). Indeed, as seen in Fig.~\ref{fig:5}, the SM predictions for $\varepsilon_K$, $\Delta M_d$ and  $\Delta M_s$ turn out to be
consistent with each other and with the data for the following values
of the CKM parameters \cite{Buras:2022wpw}
{\be\label{CKMoutput}
\vcb=42.6(4)\times 10^{-3}, \quad 
\gamma=64.6(16)^\circ, \quad \beta=22.2(7)^\circ, \quad \vub=3.72(11)\times 10^{-3}\,.
\ee
The uncertainties  shown here have been computed by propagating the non-parametric errors of the four $\Delta F=2$ observables involved. More details
  can be found
  in \cite{Buras:2022wpw}. In particular the determination  
of $\gamma$ have been obtained 
by considering first the $\vcb$-independent ratio
$\Delta M_d/\Delta M_s$ from which one derives an accurate formula for $\sin\gamma$
\be\label{singamma}
\sin\gamma=\frac{0.983(1)}{\lambda}\sqrt{\frac{m_{B_s}}{m_{B_d}}}\xi
\sqrt{\frac{\Delta M_d}{\Delta M_s}}\,, \qquad \xi=\frac{F_{B_s}\sqrt{\hat{B}_{B_s}}}{F_{B_d}\sqrt{\hat{B}_{B_d}}}=1.216(16),
\ee
with the value for $\xi$ from HPQCD \cite{Dowdall:2019bea}. The advantage of
using this ratio for the determination of $\gamma$ over studying $\Delta M_s$ and $\Delta M_d$ separately
is its independence of $\vcb$ and the reduced error on $\xi$ from LQCD relative to the individual errors of
hadronic parameters in $\Delta M_s$ and $\Delta M_d$.  See Appendix~B
in \cite{Buras:2022wpw}.
}

As emphasized in \cite{Buras:2022wpw} the very good agreement between $\Delta M_s$, $\Delta M_d$ and $\varepsilon_K$
is only obtained 
with the hadronic matrix elements with $2+1+1$ flavours from the lattice HPQCD collaboration \cite{Dowdall:2019bea}. For $2+1$ flavours significant inconsistencies
within the SM were found. See Fig.~8 in \cite{Buras:2022wpw}. 
All the input parameters used by us are collected
in Table~\ref{tab:input}.

Our value for $\vcb$ is consistent with the inclusive
one from \cite{Bordone:2021oof} and $\vub$ value with the exclusive one from
FLAG \cite{Aoki:2021kgd}. {It should be noted that the determination of $\gamma$ in this manner is more
  accurate than its present determination from tree-level decays in (\ref{betagamma}).}

\begin{figure}[t!]
  \centering%
  \includegraphics[width=0.70\textwidth]{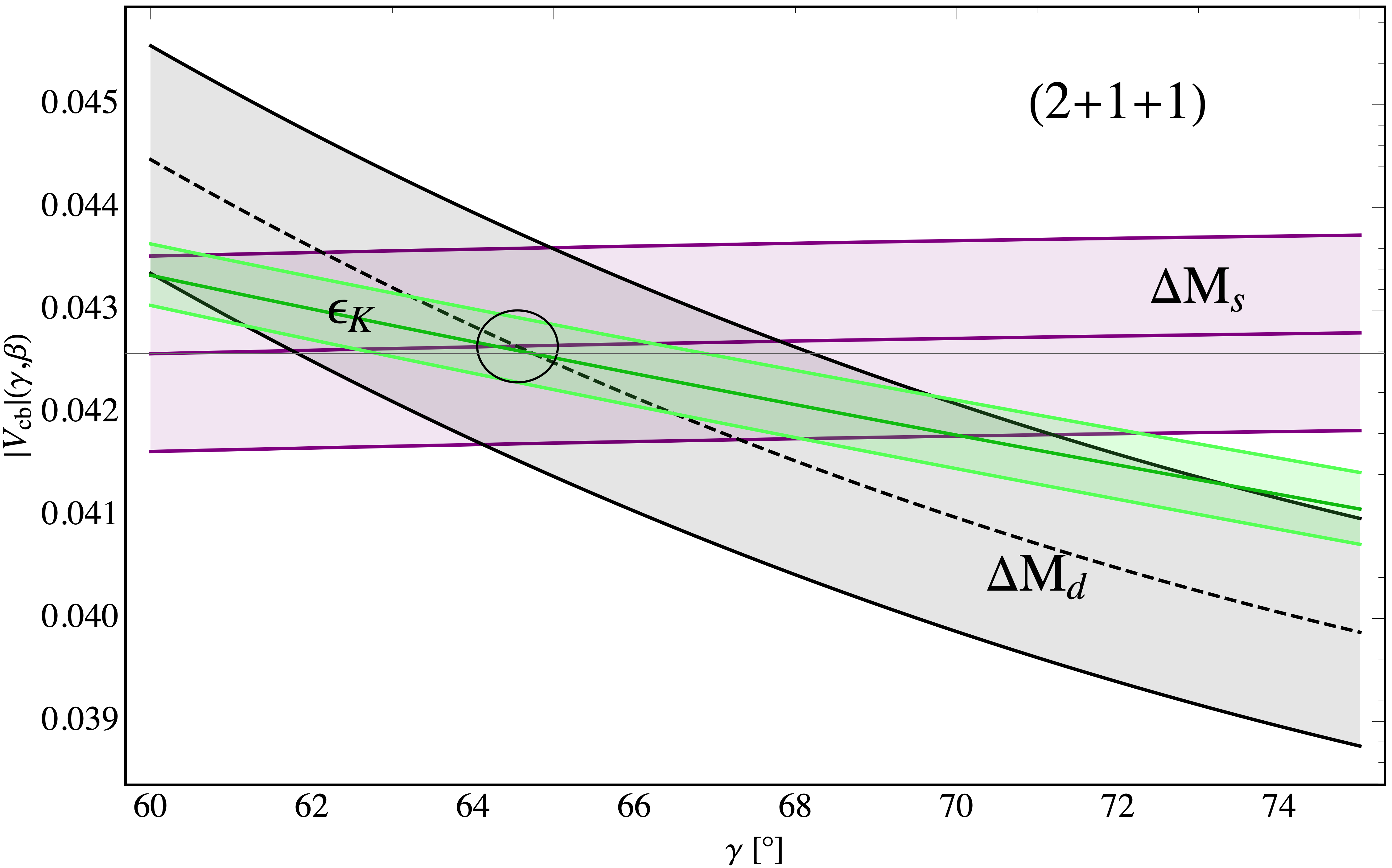}
\caption{\it {The values of $\vcb$ extracted from $\varepsilon_K$, $\Delta M_d$ and  $\Delta M_s$ as functions of $\gamma$  with  the hadronic matrix elements for $\Delta M_{s,d}$ obtained with $2+1+1$ flavours  \cite{Dowdall:2019bea}. The green band represents experimental $S_{\psi K_S}$ constraint on $\beta$. From \cite{Buras:2022wpw}.}
\label{fig:5}}
\end{figure}

In order to illustrate the action of the seven observables in the $\vcb-\gamma$
plane we show in Fig.~\ref{fig:X} the results in the SM setting all uncertainties
for transparency reasons to zero. We make the following observations.
\begin{itemize}
\item
  For fixed $\beta=22.2^\circ$, $\varepsilon_K$, $\kpn$ and $\klpn$ are represented to an excellent approximation by the same line which is already a very good test of the SM. This is simply because as seen  in (\ref{epsilon}), (\ref{Kplus}) and (\ref{KL})   the $\gamma$ dependence in the three observables is practically the same, the fact pointed out first in  \cite{Buchalla:1994tr} and
  strongly emphasized in \cite{Buras:2021nns}.
The dependence
on $\beta$ is different and this allows to determine within the SM  the angle $\beta$ from any pair of these observables
independently of the value of $\gamma$.
For the pair of the rare $K$ branching ratios this was pointed out in
  \cite{Buchalla:1994tr}. For the other two pairs in \cite{Buras:2021nns}.
  But the determination of $\beta$ with the help of the plot in Fig.~\ref{fig:X}
  is not useful and it is better to use the $\vcb$-independent ratios
  $R_0$, $R_{11}$, $R_{12}$  of  \cite{Buras:2021nns} with \cite{Buras:2022nrb}
\be
  {R_0=\frac{\mathcal{B}(\kpn)}{\mathcal{B}(\klpn)^{0.7}}=(2.03\pm 0.08)\times 10^{-3}\left[\frac{\sin 22.2^\circ}{\sin \beta}\right]^{1.4} 
={(2.03\pm 0.11)}\times 10^{-3}\,.
}
\label{eq:R0}
\ee
and
\be\label{R11R12}
R_{11}=\frac{\mathcal{B}(\kpn)}{|\varepsilon_K|^{0.82}},\qquad
R_{12}=\frac{\mathcal{B}(\klpn)}{|\varepsilon_K|^{1.18}}
\ee
with explicit expressions for 
$R_{11}$ and  $R_{12}$ given in (90) and (91) of  \cite{Buras:2021nns},
respectively. They all can be derived from (\ref{eKapp}), (\ref{kplusApprox})
and (\ref{k0Approx}) given in the previous section.
\item
  $\Delta M_d$ and $B_d\to\mu^+\mu^-$ are represented by a single line
  and a different line represents $\Delta M_s$ and  $B_s\to\mu^+\mu^-$.
  This is precisely the illustration of the SM relations pointed
  out long time ago in \cite{Buras:2003td}.
\end{itemize}

\begin{figure}[t!]
  \centering%
  \includegraphics[width=0.70\textwidth]{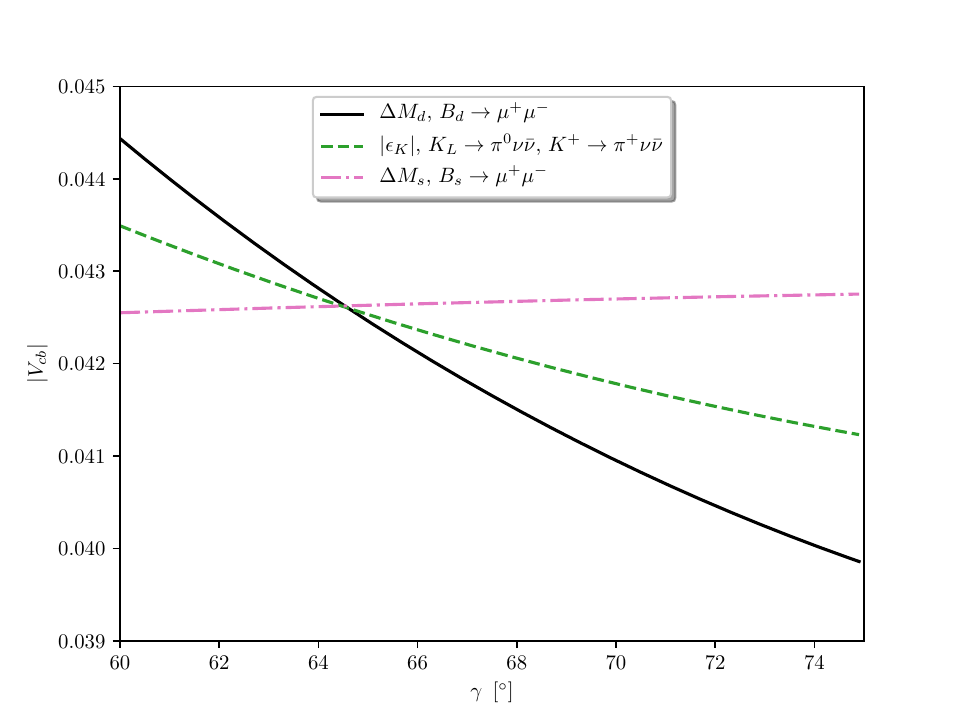}
  \caption{\it {Schematic illustration of the action of the seven observables in the $\vcb-\gamma$ plane in the context of the SM. We set $\beta=22.2^\circ$ and all uncertainties
      in (\ref{DMD1})-(\ref{Bs}) to zero.}
\label{fig:X}}
\end{figure}

\begin{figure}[t!]
  \centering%
  \includegraphics[width=0.70\textwidth]{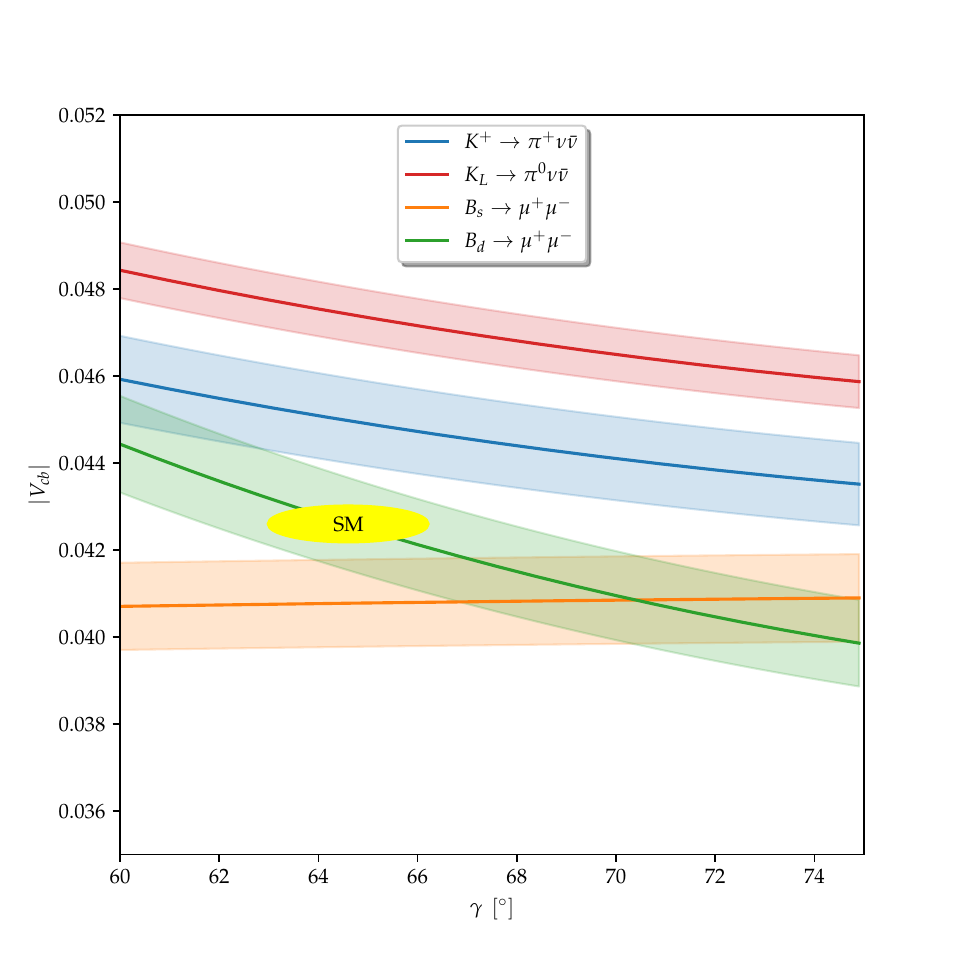}
  \caption{\it {The impact of hypothetical future measurements of 
      the branching ratios for $\kpn$, $\klpn$,  $B_d\to\mu^+\mu^-$ and $B_s\to\mu^+\mu^-$ 
      as given in (\ref{Input1}) and (\ref{Input2}) on
      the $\vcb-\gamma$ plane. All uncertainties in (\ref{DMD1})-(\ref{Bs}) are included. The yellow disc
    represents the SM as obtained in {(\ref{CKMoutput})}.}
\label{fig:Y}}
\end{figure}
While, as seen in Fig.~\ref{fig:5}, SM describes  $\varepsilon_K$,
$\Delta M_d$, $\Delta M_s$ simultaneously very well, this not need to be the case for the
four rare decays in question. This is illustrated  in Fig.~\ref{fig:Y}. To obtain these results we have
set the branching ratio for $B_s\to\mu^+\mu^-$ to
the experimental world average from LHCb, CMS and ATLAS \cite{LHCb:2021awg,CMS:2020rox,ATLAS:2020acx} in (\ref{LHCbEXP2}) but decreased its error from $8.4\%$ down to $5\%$. For the remaining branching ratios we have chosen values resulting from hypothetical future measurements that differ from the SM predictions in (\ref{BV}) and (\ref{LHCbTH}).
We kept the errors at $5\%$ as in the case of $B_s\to\mu^+\mu^-$ to exhibit the superiority  of rare $K$ decays  over rare $B$ decays as far as the determination of $\vcb$ is concerned.  We use then
\be\label{Input1}
\mathcal{B}(\kpn)= (10.00\pm 0.50)\times 10^{-11}\,,\quad
\mathcal{B}(\klpn)=(4.50\pm 0.23)\times 10^{-11}\,,
\ee
{\be\label{Input2}
\overline{\mathcal{B}}(B_{s}\to\mu^+\mu^-) = (3.45\pm 0.17)\times 10^{-9},\qquad \mathcal{B}(B_{d}\to\mu^+\mu^-)= (0.80\pm 0.04)\ \times 10^{-10}.
\ee}
While the experimental errors are futuristic, we expect that the theoretical
errors will go down with time so that the bands in Fig.~\ref{fig:Y} could
apply one day with less accurate measurements.

This plot confirms all the statements made above. The superiority of
$\klpn$ over the remaining decays is clearly seen. The blue band will be narrowed once the long distance charm contributions to $\kpn$ will be known with higher precision 
from lattice QCD calculations \cite{Christ:2019dxu} than they are known now
\cite{Isidori:2005xm}.

\begin{table}[!tb]
\center{\begin{tabular}{|l|l|}
\hline
$m_{B_s} = 5366.8(2)\mev$\hfill\cite{Zyla:2020zbs}	&  $m_{B_d}=5279.58(17)\mev$\hfill\cite{Zyla:2020zbs}\\
$\Delta M_s = 17.749(20) \,\text{ps}^{-1}$\hfill \cite{Zyla:2020zbs}	&  $\Delta M_d = 0.5065(19) \,\text{ps}^{-1}$\hfill \cite{Zyla:2020zbs}\\
{$\Delta M_K = 0.005292(9) \,\text{ps}^{-1}$}\hfill \cite{Zyla:2020zbs}	&  {$m_{K^0}=497.61(1)\mev$}\hfill \cite{Zyla:2020zbs}\\
$S_{\psi K_S}= 0.699(17)$\hfill\cite{Zyla:2020zbs}
		&  {$F_K=155.7(3)\mev$\hfill  \cite{Aoki:2019cca}}\\
	$|V_{us}|=0.2253(8)$\hfill\cite{Zyla:2020zbs} &
 $|\eps_K|= 2.228(11)\cdot 10^{-3}$\hfill\cite{Zyla:2020zbs}\\
$F_{B_s}$ = $230.3(1.3)\mev$ \hfill \cite{Aoki:2021kgd} & $F_{B_d}$ = $190.0(1.3)\mev$ \hfill \cite{Aoki:2021kgd}  \\
$F_{B_s} \sqrt{\hat B_s}=256.1(5.7) \mev$\hfill  \cite{Dowdall:2019bea}&
$F_{B_d} \sqrt{\hat B_d}=210.6(5.5) \mev$\hfill  \cite{Dowdall:2019bea}
\\
 $\hat B_s=1.232(53)$\hfill\cite{Dowdall:2019bea}        &
 $\hat B_d=1.222(61)$ \hfill\cite{Dowdall:2019bea}          
\\
{$m_t(m_t)=162.83(67)\GeV$\hfill\cite{Brod:2021hsj} }  & {$m_c(m_c)=1.279(13)\GeV$} \\
{$S_{tt}(x_t)=2.303$} & {$S_{ut}(x_c,x_t)=-1.983\times 10^{-3}$} \\
    $\eta_{tt}=0.55(2)$\hfill\cite{Brod:2019rzc} & $\eta_{ut}= 0.402(5)$\hfill\cite{Brod:2019rzc}\\
$\kappa_\varepsilon = 0.94(2)$\hfill \cite{Buras:2010pza}	&
$\eta_B=0.55(1)$\hfill\cite{Buras:1990fn,Urban:1997gw}\\
$\tau_{B_s}= 1.515(4)\,\text{ps}$\hfill\cite{Amhis:2016xyh} & $\tau_{B_d}= 1.519(4)\,\text{ps}$\hfill\cite{Amhis:2016xyh}   
\\	       
\hline
\end{tabular}  }
\caption {\textit{Values of the experimental and theoretical
    quantities used as input parameters. For future 
updates see FLAG  \cite{Aoki:2021kgd}, PDG \cite{Zyla:2020zbs}  and HFLAV  \cite{Aoki:2019cca}. 
}}
\label{tab:input}
\end{table}

\section{Conclusions and Outlook}\label{sec:4}
In the present paper we have emphasized, resurrecting by now almost thirty years old ideas of \cite{Buras:1994rj,Buchalla:1994tr,Buchalla:1996fp}, that the
rare $K$ and $B$ decay branching ratios, being subject to small hadronic uncertainties,  could soon give us a powerful tool  to determine
the CKM parameters, in particular the controversial parameter $\vcb$. They could
also provide a useful insight in the value of $\gamma$ beyond its tree-level determinations. In this context we have proposed to
monitor future progress on the determination on $\vcb$ and $\gamma$ in the
$\vcb-\gamma$ plane rather then in the  $(\bar\varrho,\bar\eta)$ plane
used in the context of the common UT-fits. We also reemphasized in this context
the important role of $\Delta M_s$, $\Delta M_d$ and $\varepsilon_K$.
To this end we derived seven expressions for $\vcb$ by means of which this CKM
element can be determined. First in the case of
\be\label{Delta2}
\Delta M_d,\qquad \Delta M_s, \qquad |\varepsilon_K|,
\ee
with the relevant expression for $\vcb$ as a function of $\gamma$, $\beta$
and the observable involved given in (\ref{DMD1}), (\ref{DMS1}) and
(\ref{epsilon}), respectively. The corresponding four formulae for $\vcb$
from 
 rare decays 
\be
\kpn,\qquad \klpn,\qquad  B_d\to\mu^+\mu^-, \qquad  B_s\to\mu^+\mu^-,
\ee
are given in (\ref{Kplus}), (\ref{KL}), 
(\ref{Bd}) and (\ref{Bs}), respectively.

The very good consistency of the three observables (\ref{Delta2}) with each other within the SM
allowed, after the imposition of the  $S_{\psi K_S}$ constraint, a satisfactory
determination of the four CKM parameters in  \cite{Buras:2022wpw} and
given in (\ref{CKMoutput}).
In the present paper this agreement is seen in the  plot 
in Fig.~\ref{fig:5}).

Possibl In the $\vcb-\gamma$ plane
it will be signaled by the inconsistency with the SM yellow disc in
Fig.~\ref{fig:Y} on which the SM prediction for this decay in  (\ref{LHCbTH}) is based.

The data on the rare decay branching ratios
allow still for significant NP contributions and inconsistencies between
various determinations of $\vcb$ as a function of $\gamma$ from different
decays  could be found one day. We illustrated it in Fig.~\ref{fig:Y}
with the SM prediction represented by a yellow disc.
In this context a precise measurement of $\gamma$ by the LHCb and Belle II collaborations and the improvements on $\beta$ will allow a very precise determination of $\vcb$
within the SM, first with the help of precisely measured $|\varepsilon_K|$ and later $\kpn$ and $\klpn$.

 Simultaneously the very accurate $\vcb$ independent SM predictions
for rare decay branching ratios found in \cite{Buras:2021nns,Buras:2022wpw}
and recalled here in (\ref{BV}) and (\ref{LHCbTH}) will play,
in case of  inconsistencies in the $\vcb-\gamma$ plane, an 
important role in the identification of a particular  NP at work. The 16 $\vcb$-independent ratios of various flavour observables
 derived in \cite{Buras:2021nns} will also be useful in this context.

We are looking forward to new data from LHCb, NA62, KOTO and Belle II collaborations as well as to improved hadronic matrix elements from LQCD which
will allow one to use this strategy and the ones outlined
in \cite{Buras:2021nns,Buras:2022wpw}  more efficiently than it is possible now.

{\bf Acknowledgements}
I would like to thank Elena Venturini for the most pleasant and  efficient
collaboration leading to \cite{Buras:2021nns,Buras:2022nrb,Buras:2022wpw}
that had clearly an important impact on the present paper. In particular
I would also like to thank her for continuous discussions on classical music,
first of all  the one of Bach, Beethoven, Brahms, Mozart, Chopin, Rachmaninow, Schumann, Schubert, Vivaldi and Tschajkowski,  but  recently also of Alma Deutscher, the {\em Mozarta} of the 21st century. The invaluable help from Marcin Chrzaszcz in constructing the plots in Figs.~\ref{fig:X} and \ref{fig:Y} is highly appreciated.
\noindent
Financial support from the Excellence Cluster ORIGINS,
funded by the Deutsche Forschungsgemeinschaft (DFG, German Research Foundation), 
Excellence Strategy, EXC-2094, 390783311 is acknowledged.

\renewcommand{\refname}{R\lowercase{eferences}}

\addcontentsline{toc}{section}{References}

\bibliographystyle{JHEP}

\small

\bibliography{Bookallrefs}

\end{document}